# ROBUST PHASE BEHAVIOR OF MODEL TRANSIENT NETWORKS

Mohammed FILALI*§ , Mohamed Jamil OUAZZANI§, Eric MICHEL*, Raymond AZNAR*,

Grégoire PORTE*, Jacqueline APPELL*[+],

* Groupe de Dynamique des Phases Condensées  UMR5581 CNRS-Université Montpellier II, CC 26 ,

34095 Montpellier cedex5, FRANCE and  § Faculté des Sciences Laboratoire de Physique du Solide,

Dharmehraz BP1796 Atlas , FES, Morocco

**ABSTRACT**

In order to study the viscoelastic properties of certain complex fluids which are described in terms of a multiconnected transient network we have developed a convenient model system composed of microemulsion droplets linked by telechelic polymers. The phase behavior of such systems has two characteristic features:  a large monophasic region which consists of two sub-regions (a fluid sol phase and a viscoelastic gel phase) separated by a percolation line and a two phase region at low volume fraction with separation into a dilute sol phase and a concentrated gel phase. From the plausible origin of these features we expect them to be very similar in different systems. We describe here the phase behavior of four different systems we prepared in order to vary the time scale of the dynamical response of the transient network; they consist of the combination of two oil(decane) in water microemulsions differing by the stabilizing surfactant monolayer (Cetyl pyridinium chloride/ octanol or TX100/TX35) and of two telechelic polymers which are end-grafted poly (ethylene oxide) chains, differing by the end-grafted hydrophobic aliphatic chains ($C_{12}H_{25}$ or $C_{18}H_{37}$). We first summarize  the characterization of the structure of the four systems by small angle neutron scattering: the size of the microemulsion droplets is found to be constant in a given system upon addition of a telechelic polymer. In the CPCl systems we find a mean radius of the  microemulsion droplets = 62± 1Å and a very narrow size distribution and in the TX systems we find a mean radius = 84± 2Å and a somewhat  larger size distribution. We can then calculate precisely the number of polymers per microemulsion droplet and compare the phase behavior of the four systems in consistent units. As expected we find very similar phase behavior in the four systems.

+ to whom correspondence should be adressed. E-mail appell@gdpc.univ-montp2.fr

**INTRODUCTION**

The viscoelastic properties [1] of many complex fluids are described in terms of a transient network characteristic of physical gels [2]. To investigate these properties, a model system has been developed in our group following the pionneering work in references [3-5]. It consists of droplets of an oil in water (O/W) microemulsion and of telechelic polymers: a water soluble poly(ethylene oxide) chain modified by grafting an aliphatic chain at both ends. These end-chains stick into the hydrophobic core of the microemulsion droplets and can either decorate or bridge the droplets. We have shown in a previous paper that bridging indeed occurs in such a system which then becomes a multiconnected transient network [6].

The main advantage of this pseudo-ternary model system is that we can monitor separately different parameters playing a role in the properties of the transient network, namely the radius of the droplets by adjusting the composition of the microemulsion, their average distance depending on the volume fraction of the droplets and the number of telechelic polymers added per droplet. We can thus tailor the network by adjusting the size and number density of knots and the number of linking chains. This is a net advantage over similar networks formed in simple binary solutions of associative polymers where all these parameters are dependent on the chosen polymer and on the particular concentration used. Furthermore a practical advantage of such a system is that the scattering of neutron (or light) is mainly due to the droplets, the polymer contribution being negligible. Information is then easily obtained on the shape and size of the droplets and on the interactions introduced between them by studying the small angle neutron scattering patterns [4-7].

In the phase behavior of these systems two interesting features are observed. First a phase separation occurs at low volume fraction between a dilute solution and a concentrated solution and second the large monophasic range can be divided into two sub-ranges -a fluid sol and a viscoelastic transient gel separated by a percolation line. From the origin of the phase separation and of the percolation line discussed below, we can predict a large similarity of these features in different systems and this is indeed what is observed and described in this paper.

At low volume fraction $\Phi$ and at moderate concentration of polymers a phase separation between a very dilute sol and a concentrated gel is observed. This phase separation is an associative phase separation [8] brought about by an effective attractive interaction between the droplets. This effective attraction originates in the possibility for a telechelic polymer to link two microemulsion droplets. The change in free energy (i.e. the adhesion energy of a sticker) is identical when the hydrophobic extremities experience an apolar environment no matter whether they are in the same or in two different droplets [9]. Why then, do we observe this effective attraction ? When the droplets are far apart, the chains are too short to bridge them: a chain having one of its stickers in one given droplet is forced to loop so that its second sticker adsorbs onto the same droplet. When the droplets are close enough to one another (at a distance of order $R_g$ the radius of gyration of the telechelic polymer chain) loop conformations are still allowed and, in addition, bridging conformations are now accessible. So the conformational entropy of the chain is larger when the droplets are at the right distance for bridging. This very simple argument was first proposed by Witten [10] : assuming that the numbers of loop and bridge conformations are roughly equal, he derived the free energy change in bringing the droplets close to each other: $-k_B T \ln(2)$ per telechelic chain. More refined calculations were reported [11,12] later on, for the effective bridging interaction between flat surfaces in the different regimes (mushroom and brush) for the area density of telechelic chains: a net attraction is again found but with a magnitude somewhat lower. The effect of the bridging attraction onto the phase behavior of solutions of associating polymers is further analyzed theoretically in [13]. Experimentally this phase separation has been observed in binary systems [14] as well as in ternary systems [4-7]. We have discussed at length [6] this effective interaction per se and as the driving force for the phase separation.

The second feature in the phase behavior is the evolution of the system from a sol phase which flows easily to a gel-like phase which displays viscoelastic properties. In the gel phase, the telechelic polymers bridge the droplets forming a multiconnected network where the droplets are the knots and the polymers the links. At the onset of this regime corresponds a percolation threshold where one connected cluster spans the entire sample[15]. The percolation line is the locus of the percolation thresholds . In this picture, the percolation line depends on the connectivity of the network but not on the adhesion energy of the sticker although this adhesion energy is an important parameter of the viscoelastic properties such as the relaxation time.

In order to check that the phase separation and the percolation line are, as expected from the description above, independent of the adhesion energy of the stickers we want to compare the phase behavior of different systems. We use four different mixed systems formed from two different microemulsions (differing in the constituents of the surfactant layer) and two telechelic polymers (differing in the length of the aliphatic chains). In order to make valuable comparisons, the networks must be well characterized and in particular their connectivity quantitatively measured. In order to calculate the number of polymers by droplet and thus the connectivity, we need an accurate determination of the size of the microemulsion droplets. This can be obtained using small angle neutron scattering data. We report here on these measurements. We will then be able to discuss the phase behavior of the systems and check their similarity. Furthermore this will be useful in comparing properly the dynamical properties of such networks, using different systems we can vary the time scale of the dynamical response of the transient network [15-18].

**EXPERIMENTAL**

**Materials**:

Cetyl pyridinium chloride $[H_3C\text{-}(CH_2)_{15}]\text{-}C_5H_5N^+\ Cl^-$ (CPCl) from Fluka is purified by successive recristallization in water and in acetone, octanol $[H_3C\text{-}(CH_2)_7]\text{-}OH$ and decane $[H_3C\text{-}(CH_2)_8\ CH_3]$ from Fluka are used as received. The non-ionic surfactants TX100 and TX35 are purchased from Sigma Chemicals and used as received

The poly (ethylene-oxide) have been hydrophobically modified and purified in the laboratory using the method described in [19, 20]. The molecular weight of the starting products is determined by size-exclusion chromatography. The hydrophobically modified poly(ethylene-oxide) contains an isocyanate group between the alkyl chain and the ethylene-oxide chain. We assume this isocyanate group belongs to the hydrophilic part of the copolymer. Two telechelic polymers have been prepared: poly (ethylene-oxide) PEO-C12 with a $C_{12}\ H_{25}$ aliphatic chain grafted at each extremity and PEO-C18 with a $C_{18}\ H_{37}$ aliphatic chain grafted at each extremity. After modification, the degree of substitution of the hydroxyl groups was determined by NMR using the method described in [21]. The degree of substitution is found to be equal or larger than 98% .

All samples are prepared by weight. For the TX systems, they are prepared in triply distilled water or in deuterated water (from Solvants Documentation Synthese Co) used as received. For the CPCl systems, they are prepared in 0.2M-NaCl brine or deuterated brine. The samples are characterized by their volume fraction Φ of aliphatic chains (from decane, surfactant layer and PEO-C12 or PEO-C18) which form the hydrophobic cores of the microemulsion droplets, and by the number r of $C_{12}$ or $C_{18}$ chains per droplet. All the parameters necessary to calculate Φ and r from the sample composition are summarized in Table 1.

### Preparation of the microemulsions :

The microemulsions[22] are here thermodynamically stable dispersions in water of oil droplets surrounded by a surfactant film: O/W microemulsions. The spontaneous radius of curvature of the surfactant film is adjusted by varying its composition. The composition of the two surfactant films used are given in table 2 together with the ratio in weight of decane to surfactant film. This ratio is chosen in order to be close but slightly below the emulsification failure limit. The line of emulsification failure is the limit above which the microemulsion droplets are saturated with oil and coexist with excess oil. On this line the microemulsion droplets have a radius corresponding to the spontaneous curvature radius of the surfactant film[23]. Under such conditions it is now well established that the droplets of microemulsion are spheres of a well-defined radius[24] and that they can be diluted over a large concentration range[25,26]. We did find that the microemulsions can be diluted over the range of ~1 to 15 weight %. In this range the microemulsion droplets are fairly monodisperse spheres as described below.

**Preparation of the microemulsion droplets plus telechelic polymers:**

The samples are prepared by weight. Their overall composition is determined so as to obtain a constant volume fraction $\Phi$ of the hydrophobic parts (HC) of the droplets (which consist of the hydrophobic parts of the microemulsion constituents plus the alkyl chains of the PEO-C12 or PEO-C18) while increasing progressively the number of adsorbed alkyl chains. This is achieved by replacing a small amount of the surfactant film by the appropriate amount of modified PEO. To calculate the number $\underline{r}$ of $C_{12}$ or $C_{18}$ chains per droplet, we assume that the radius of the spherical droplet does not change with increasing substitution of the surfactant by the copolymers; we showed this previously for one of the system [6] and further evidence is given below. The precision on the value of $\underline{r}$ depends on the accurate determination of the number of microemulsion droplets per unit volume and thus on the accurate determination of their size.

**Observation of the phase behavior of the samples**

The samples prepared as described above are thoroughly shaken to insure homogenization and then kept at the temperature of observation, here T=20°C or 25°C, in a thermostated water bath for several days before visual examination. When a phase separation is observed the samples are rehomogenized and set back to rest for a couple of days to confirm the observations.

The percolation line, which separates the monophasic region in two sub-regions one of fluid sol phase and one of viscoelastic transient gel phase, can in principle be determined experimentally from the results of rheological measurements as described previously [15] for the TX/PEO-C18 system. We will discuss below why it can, in fact, be determined reliably only in the TX or CPCl/PEO-C18 systems

**Small angle neutron scattering: SANS Measurements**

They have been performed at LLB-Saclay on the spectrometer PACE. The range of scattering vectors covered is 0.004 Å$^{-1}$ < q <0.16 Å$^{-1}$ . The temperature is T= 20°C. The scattering data are treated according to standard procedures. They are put on an absolute scale by using water as standard. And we obtain intensities in absolute units (cm$^{-1}$) with an accuracy better than 10%. To simulate correctly the experimental spectra all the model spectra are convoluted by the instrumental response function taking into account the uncertainty on the neutrons wavelength and the angular definition [27].

To gain in accuracy in the determination of the size of the microemulsion droplets we made experiments under two different contrast conditions. The first configuration is the classical one: the droplets from hydrogenated constituents are in solution in deuterated water or brine, in what follows it is referred to as sphere contrast. In the second configuration the droplets of deuterated decane are surrounded by an hydrogenated surfactant film and dispersed in deuterated water or brine. Deuterated decane and deuterated water have almost equal scattering length density (6.6 10$^{10}$ cm$^{-2}$ and 6.4 10$^{10}$ cm$^{-2}$) so that a droplet is viewed as a spherical shell formed by the surfactant layer; in what follows it is referred to as shell contrast.

The small angle neutron scattering from colloidal solutions provides information on their structure [28, 29]. If the colloidal aggregates can be assumed to be spherical or if, at least on average, the interaction potential between them has spherical symmetry, one can write the scattered intensity I (cm$^{-1}$) in the form:

$$I(q) = \Phi \, v (\Delta\rho)^2 P(q) S(q) = A P(q) S(q)$$
$$\text{with } A = \Phi \, v (\Delta\rho)^2 \quad (1)$$

where q (Å$^{-1}$) is the scattering vector; $\Phi$ is the volume fraction of aggregates (sphere or shell); v (cm$^3$) the dry volume of the aggregates (sphere or shell) and $\Delta\rho$ (cm$^{-2}$) the contrast i.e. the difference in the scattering length density of the aggregates and of the solvent. P(q) is the form factor of the colloidal aggregates and $P(q\rightarrow 0)=1$. S(q) is the structure factor which reflects interactions between the aggregates, at large values of q, S( q large )->1. We will focus on the range of q where S~ 1

The model spectra have been calculated using the appropriate form for P(q) in each case. In all cases the polydispersity of the droplets size is described by a gaussian distribution of the radius of the droplets with a mean radius $\bar{R}$ and a standard deviation $\Delta R$; assuming that the droplets are spherical. Then we write:

$$I(q) = \Phi(\Delta\rho)^2 \int \frac{v}{\sqrt{2\pi}\Delta R} P(q,R) \cdot e^{-\frac{(R-\bar{R})^2}{2(\Delta R)^2}} dR \qquad (2)$$

In the sphere contrast

$$P(q) = \left[\frac{3[(\sin qR) - qR\cos qR]}{(qR)^3}\right]^2 \qquad (3)$$

with R the radius of the sphere; in the limit of large q, we will use the Porod representation ($q^4 I(q)$ as a function of q) which amplifies the oscillations of P(q). If I(q) is given by (3) then, in the Porod representation, $q\bar{R}$ =2.73 and 6.12 for the first and second maximum and $q\bar{R}$ = 0, 4.49 for the first and second minimum [30].

In the shell contrast

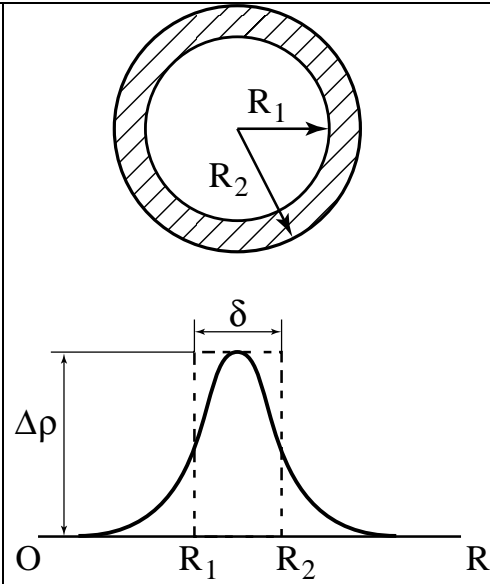

Figure 1:

Illustration of the shell contrast for the microemulsion droplet. The contrast profile is shown as a dotted line for sharp boundaries and as a full line for diffuse boundaries (see text)) We define $R_m = (R_1+R_2)/2$ and $\delta = R_2-R_1$.

For a model of concentric shells with sharp boundaries (cf figure 1) . We write the form factor $P(q)$ as a function of $R_m = (R_1+R_2)/2$ and $\delta = R_2-R_1$

$$P_{shell}(q) = \frac{9}{\delta^2 (3R_m^2 + \frac{\delta^2}{4})^2 q^6} \left[2qR_m \sin qR_m \sin q\delta/2 + 2 \sin q\delta/2 \cos qR_m - q\delta \cos qR_m \cos q\delta/2\right]^2 \quad (4)$$

and $v_{shell} = \frac{4\pi}{3} \delta (3R_m^2 + \frac{\delta^2}{4})$

The size distribution is introduced as above assuming a constant $\delta$ and a gaussian distribution of $R_m$.

If $\delta \ll R_m$ and $q\delta <1$ (4) reduces to $P_{shell}(q) = \sin^2 (qR_m)/(qR_m)^2$. We will use a $q^2 I(q)$ representation which amplifies the oscillations of $P(q)$; in a first approximation the extrema are those of the function $\sin^2 (qR_m)$.

**RESULTS and DISCUSSION.**

In what follows we indicate the values for $\bar{R}$ the mean radius, $\Delta R$ the standard deviation of the distribution of size and in the shell contrast $\bar{R}_m$ the mean radius, $\delta$ the thickness of the shell and $\Delta R$ obtained for the best adjustment of the computed spectra to the experimental results, together with an estimation of the uncertainty on these values.

---

### The CPCl/Octanol/decane bare microemulsion

We have discussed previously [6] at length the results obtained for this microemulsion under sphere contrast. In figure 2 the spectra, obtained for the microemulsion at different $\Phi$, are displayed in the Porod representation together with the spectra calculated as explained above with $\bar{R}= 62\pm 1$Å and a standard deviation of the distribution of size $\Delta R= 6.5 \pm 1$Å. All spectra are identical at high q , the differences at small q are due to the structure factor S(q) which can be taken into account assuming an interaction between droplets which is the sum of a van der Waals attraction and of a screened coulombic repulsion as described in [6].

The results obtained in the shell contrast are shown in figure 3. Here again all spectra are identical at high q and are well described by the spectra calculated using relations (2) and (4) with $\bar{R}_m = 54.5\pm 1$Å, $\delta = 13 \pm 1$ Å and $\Delta R = 6.5 \pm 1$ Å . These values are in excellent agreement with those obtained in shell contrast. Consistently, the radius of the "dry" microemulsion droplet is $\bar{R}= 62\pm 1$Å.

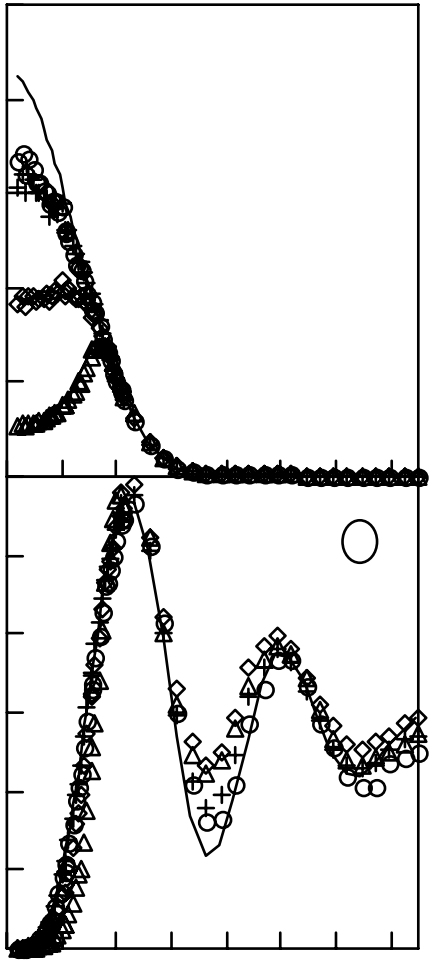

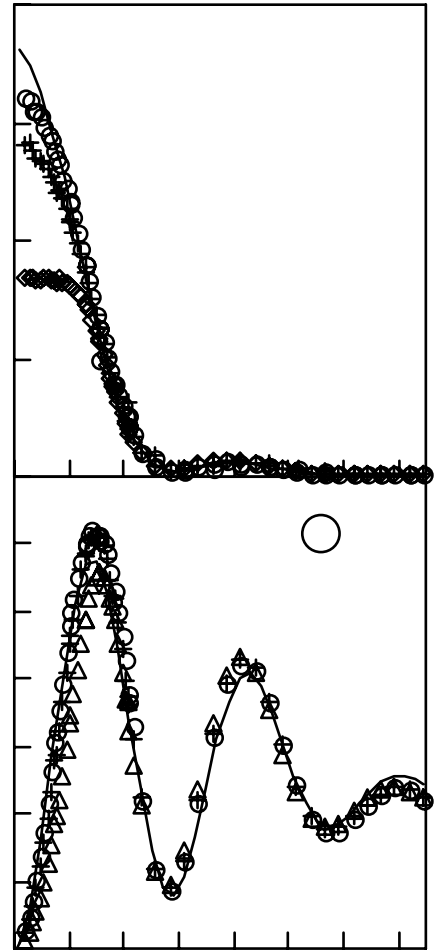

Figure 2:

Spectra for the CPCl microemulsion in sphere contrast: A: O: Φ= 0.014; +: Φ =0.028; ◊: Φ =0.07; Δ: Φ =0.137; and line = spectra computed for $\bar{R}$= 62 ± 1Å and ΔR= 6.5 ± Å . B The Porod representation amplifies the form factor oscillations.

Figure 3

Spectra for the CPCl microemulsion in shell contrast. A: The experimental spectra are normalized to a shell volume fraction = 1  O: Φ= 0.014; +: Φ =0.028 Δ: Φ =0.069 and line = spectra computed for $\bar{R}_m$= 54.5 ± 1 Å, δ= 13±1 and ΔR= 6.5 ± 1 Å. B: The same spectra in the $q^2 I(q)$ representation which amplifies the form factor oscillations.

|  |  |
|--|--|
|  |  |

## The TX100/TX35/decane bare microemulsion

The spectra obtained for the microemulsion at different volume fraction under sphere contrast are displayed in figure 4 together with the calculated spectra using relations (2) and (3) with $\bar{R}= 84\pm 2$ Å and $\Delta R =15$ Å..

Under shell contrast the experimental spectra are shown in figure 5 together with the calculated spectra. We first tried to calculate the spectra using relations (2) and (4) as above but obtained no reasonable agreement. We trace this back to the fact that the interpenetration of the surfactant and of decane on the apolar side and of water on the polar side cannot be neglected using TX100 and TX35 as surfactants so that the assumption of a shell with sharp boundaries is incorrect. For a similar microemulsion with a nonionic surfactant, Gradzielski et al [31] have derived the scattered intensity assuming a shell with diffuse boundaries. The contrast profile is described by a gaussian distribution of scattering length density so that the contrast $\rho(R)$ can be written:

$\rho(R) = \Delta\rho \exp(-(R - R_m)^2 / 2t^2)$    et  $t \sim \delta/(2\pi)^{0.5}$  where $R_m$ and $\delta$ are the mean radius and the width of the shell model with sharp boundaries as above and $\Delta\rho$ is the maximum of contrast reached at $R_m$ (cf figure 1). The intensity scattered for a shell with radius $R_m$ and volume fraction $\Phi_i$ is then given by

$$I_i(q) = \Phi_i \frac{12\pi * 10^{-4}}{\left[3\delta R_m^2 - \frac{\delta^3}{4}\right]} \Delta\rho^2 \frac{\delta^2}{q^2} \exp-(qt)^2 \left[R_m \sin qR_m + qt^2 \cos qR_m\right]^2 \qquad (5)$$

The instrumental response function and the size distribution (assuming a constant $\delta$ and a gaussian distribution of $R_m$) are introduced as above. The resulting spectra calculated with $\bar{R}_m = 80\pm 4$ Å, $\delta = 17\pm 2$ Å and $\Delta R = 15 \pm 1$ Å can be seen to be a good adjustment of the experimental patterns in figure 5. We note that the agreement between the results obtained in the two situations of contrast is not as good as in the case of the CPCl microemulsion, this is certainly due to the very diffuse interfaces which are more difficult to introduce properly into a model spectra. Furthermore the size distribution of droplets is broader p= $\Delta R/\bar{R}$ = 0.18 compared to p = 0.10 in the CPCl microemulsion so that the assumption of a constant thickness $\delta$ for the surfactant layer is certainly less appropriate here. Consistently with these results, the "dry" microemulsion droplet has a mean radius $\bar{R}$= 84± 2Å.

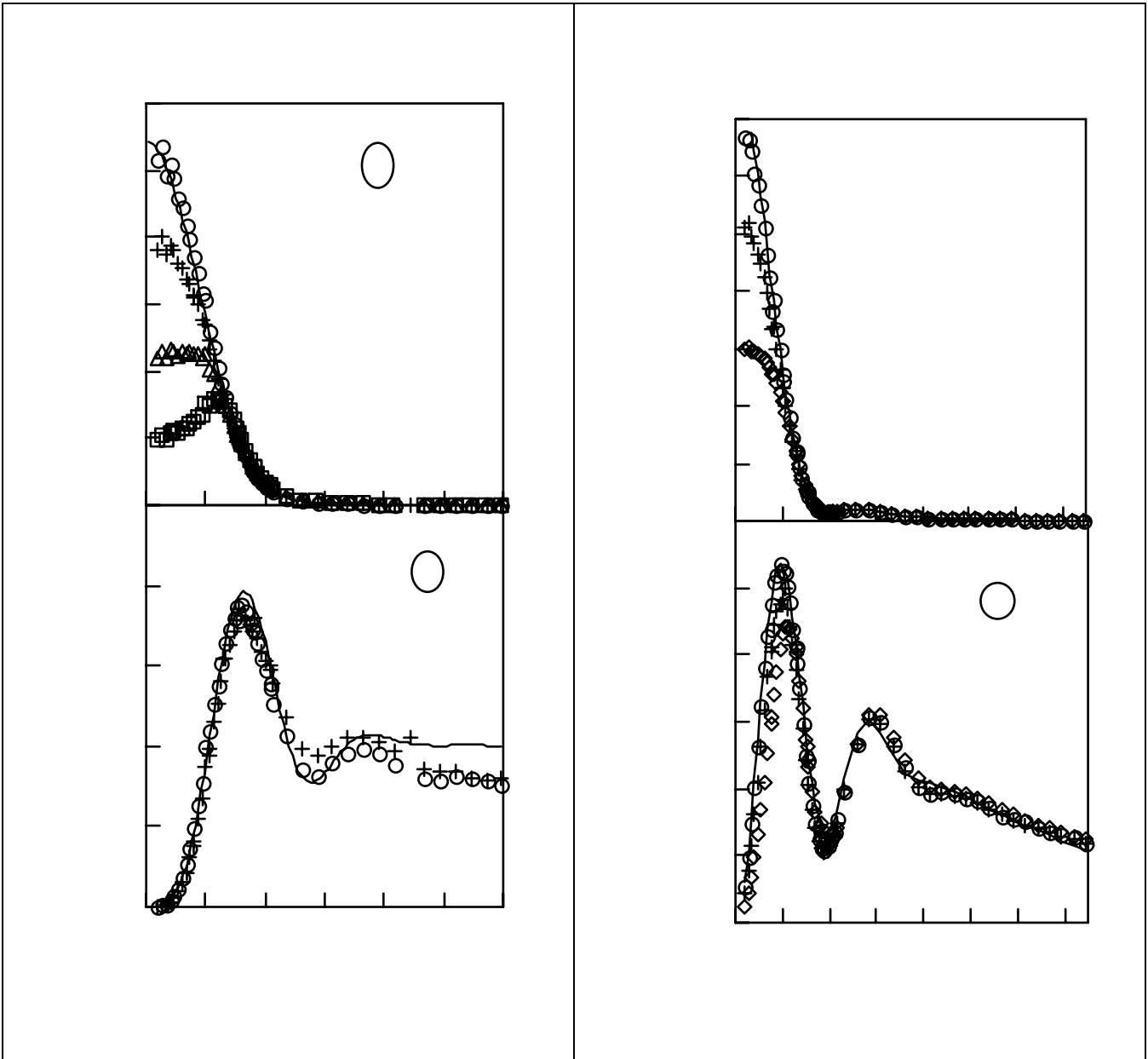

Figure 4: Spectra for the TX microemulsion in sphere contrast. A : O Φ=0.014; + :Φ =0.028; Δ: Φ =0.069;  =0.136 and line = spectra computed for $\bar{R}$=84± 2 Å and

Figure 5 : Shell pattern for the TX microemulsion: the experimental spectra, measured for increasing volume fraction of microemulsion are normalized to a shell volume

| | |
|---|---|
| ΔR =15± 1Å<br>B: The Porod representation amplifies the form factor. We plot only two of the experimental spectra at O :Φ =0.014; and + :Φ =0.028 for clarity. | fraction = 1.<br>A O :Φ = 0.014, +: Φ = 0.028; ◊ :Φ =0.07 The solid line is the computed spectra (see text) with a diffuse shell contrast given by (5) with $\bar{R}_m = 80\pm 2$ Å, $\delta=17\pm 2$ Å and ΔR= 15±1 Å. B: the same spectra in the representation $q^2I(q)$ versus q which amplifies the oscillations of the form factor. |

### **Addition of telechelic polymers to the microemulsions.**

**Calculation of the number of polymers per droplets.**

As already stated, in order to compare the four systems studied, we must determine correctly the number of polymers per droplet. Having measured precisely the size of the microemulsion droplets we have checked that this size does not change upon incorporation of telechelic polymers (see ref(6) and below). For each sample we can thus calculate the number of droplets per unit volume and the average number of stickers per droplet is given by:

$$r = \frac{v_{drop}}{\Phi_{drop}} \frac{[m]}{M} \frac{N_a}{2} \quad \text{with} \quad v_{drop} = \frac{4\pi \bar{R}^3}{3} \tag{6}$$

[m] is the mass concentration and M the molar mass of the polymer and $N_a$ Avogadro's number.

In our previous publications we had made this calculation but the determination of the radius was less precise and we had, incorrectly, considered the radius measured was that of the hydrophobic part of the droplets alone. We now, correctly, consider that the radius corresponds to the entire "dry" droplet. Both facts leads to a correction of the r's. In ref (6,16) dealing with systems based on the CPCl microemulsion the r's must be multiplied by 0.9 while in ref (15,17,18) dealing with systems based on the TX microemulsion the r's must be multiplied by 0.74.

**The droplets don't change size.**

We have shown previously [6] that the droplets of the CPCl microemulsion remained identical upon addition of relatively large amount of PEO-C12 telechelic polymers (r <= 36). This is also the case when PEO-C18 is added to the CPCl microemulsion as illustrated in figure 6 or when PEO-C12 is added to the TX microemulsion as shown in figure 7.

**Interactions between the droplets**

In the bare microemulsions, we have stressed above that the lower and lower intensity scattered at small q's, when the volume fraction increases, is the signature of a repulsive interaction between droplets. In the CPCl microemulsion this repulsion is due to a screened coulombic interaction while in the TX microemulsion, where it is less pronounced, it is due to the repulsion between the corona of ethylene oxide small chains surrounding the drops and avoiding interpenetration.

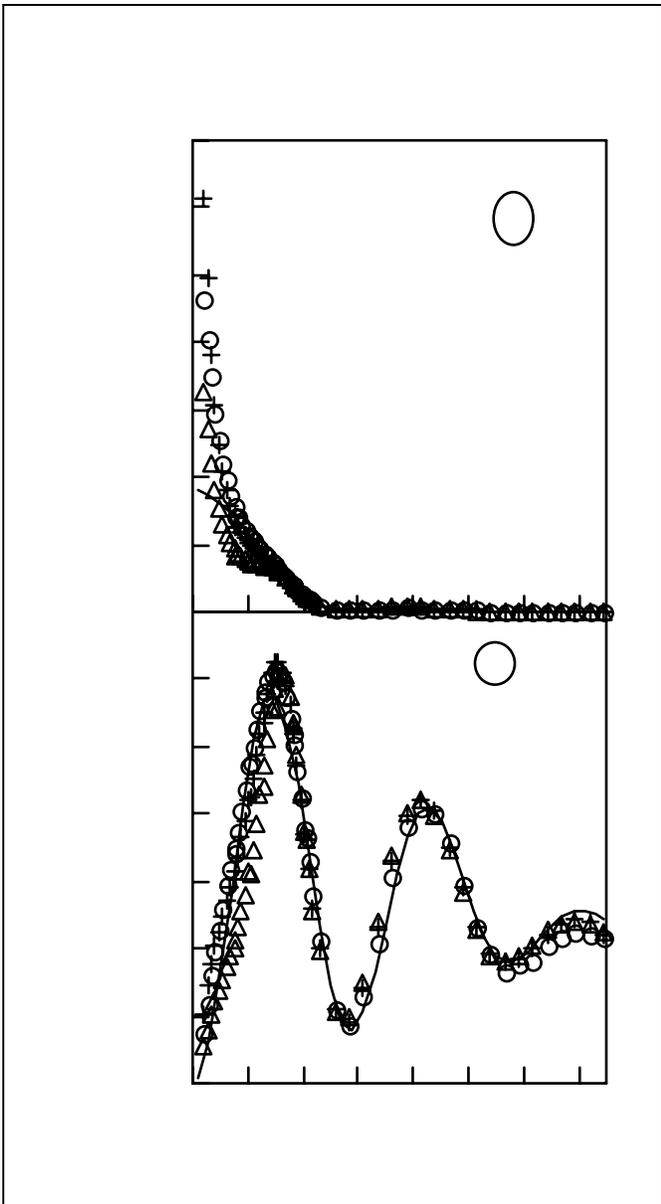 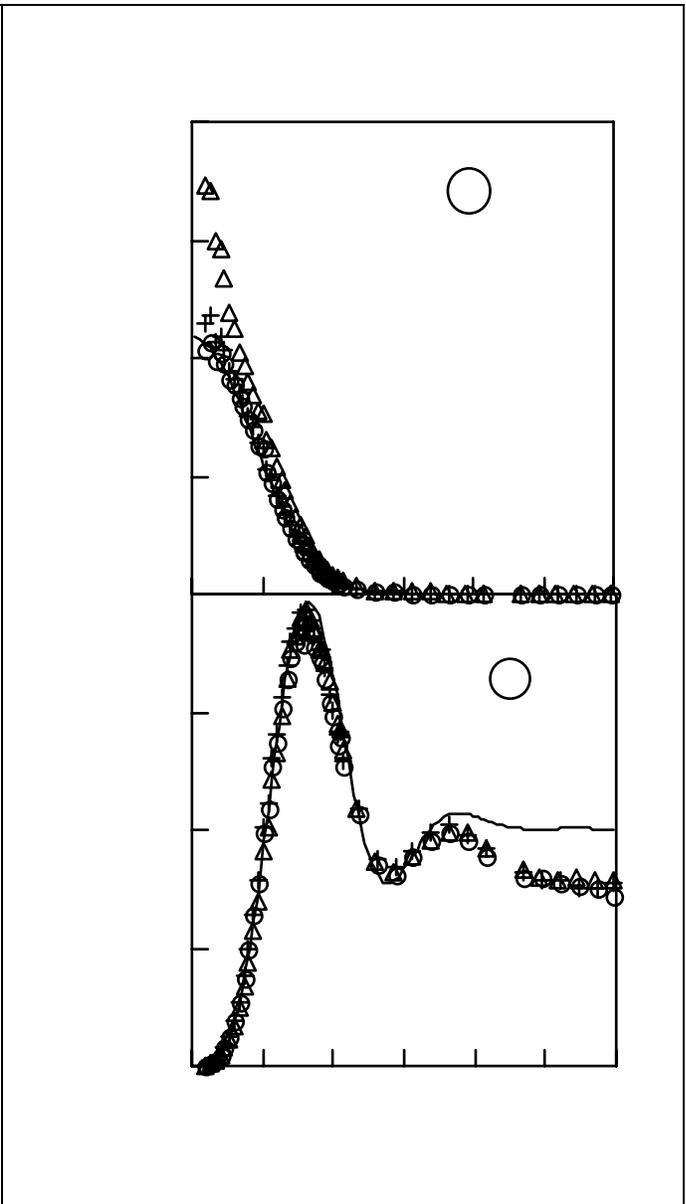

Figure 6 : Spectra for the CPCl microemulsion + PEO-C18 r= 7.2 in shell contrast. A: O: $\Phi$= 0.014; +: $\Phi$ =0.028 $\Delta$: $\Phi$ =0.07 and line = spectra computed for $\bar{R}_m$= 54.5 ± 1 Å, $\delta$= 13±1 and $\Delta R$= 6.5 ± 1Å note the large increase of I(q) at small q which is the signature of an effective attractive interaction between the droplets (see text). B The $q^2 I(q)$ representation amplifies the form factor oscillations. The identity of the spectra in the high q range indicates the droplets of microemulsion remain identical upon addition of PEO-2M.

Figure 7 : Spectra for the TX microemulsion + PEO-C12 in sphere contrast. $\Phi$= 0.014 with O: r= 0; +: r =1.8 $\Delta$: r =5.3 and line = spectra computed for $\bar{R}$ = 84 ± 1 Å and $\Delta R$= 15 ± 1 Å. A: note the large increase of I(q) at small q which is the signature of an effective attractive interaction between the droplets introduced by PEO-C12 (see text). B The $q^4 I(q)$ representation amplifies the form factor oscillations and the identity of the spectra in the high q range indicates the droplets of microemulsion remain identical upon addition of PEO-C12.

Upon inspection of figures 6 and 7, a noticeable difference is apparent: the intensity at q-> 0 increases with the decreasing volume fraction of droplets and a constant number of polymers as shown in figure 6 or with an increasing number of polymers at a given volume fraction as shown in figure 7. This indicates that addition of telechelic polymers to the microemulsion introduces an effective attractive interaction between the droplets. We have described, in the introduction above, the origin of this interaction. The attractive component can be pictured as due to the longer time the droplets spend close to one another in the low volume fraction range because of the, then possible, bridging which leads to more accessible configurations for the PEO chain and thus to an increased entropy. At higher volume fraction, when the distance between droplets become equal or smaller than the end-to-end distance of the polymer chain, bridging can occur without problem and the net effective interaction is repulsive, the polymer chains add to the repulsion because they resist interpenetration and swell in water or brine. This effective attractive interaction is found in all four systems as illustrated previously [6] and in figures 6 and 7.

**Phase behavior**

The phase behavior of the four systems is displayed in figures 8 and 9:

Over a large range of $\Phi$ and r the samples are monophasic but this monophasic range can be splitted in two sub-ranges according to the rheological properties of the sample. At low $\Phi$ and r the samples are a sol: they flow easily. At higher $\Phi$ and r the samples are a gel, they form a transient network and are viscoelastic. The line separating this two sub-ranges is a percolation line of the network as described above in the introduction and discussed in [15] for the TX / PEO-C18 system.

The determination of the percolation threshold has been described at length in [15]. The stress relaxation curve G(t) is measured after a step strain and is fitted to a stretch exponential yielding the instantaneous elastic modulus G(0) and the relaxation time $\tau_R$ as illustrated in figure 10 for a TX / PEO-C18 sample. G(0) and $\tau_R$ both follow a power law in $(r-r_P)$ as expected in a percolation behavior. As an illustration, G(0) and $\tau_R$ are plotted, in figure 11, as a function of r, allowing for the determination of the percolation threshold $r_p$ and of the exponent of the power law. The percolation line is indicated in figure 8B and 9B for the CPCl /PEO-C18 and TX/PEO-C18 systems. As expected from its description, it is indeed very similar going from one to the other system. As described above, the determination of the percolation threshold rests on a change in the evolution of the rheological properties with $\underline{r}$ in the sol phase or in the transient gel phase. This threshold is smeared by the fact that i/before the occurrence of an infinite cluster (i.e. the percolation threshold) more and more clusters of increasing size are formed and ii/ after the occurrence of the first infinite cluster, a large part of the droplets are still free or belongs to smaller clusters. However in the two C18-systems the change in rheological properties is rather abrupt: the diffusion of non-connected drops and small clusters in the sol phase leads to a fast relaxation of the stress compared to the much slower relaxation when rupture of the links in the transient gel is the rate limiting mechanism. In the TX/PEO-C12 and CPCL/PEO-C12 we are convinced that the percolation line exists but in contrast to the two other systems the change in rheological properties is very smooth and does not allow to point out the percolation threshold. This can be understood if we recall that the links break by extraction of stickers from the droplets so that the time scale on which the rupture occurs is directly related to the energy of adhesion of the sticker[9] and can be estimated to be smaller by more than three order of magnitude when switching from C18 to C12 stickers. Thus in the C12 systems it probably becomes comparable to the time scale of diffusion of non-connected drops so that the evolution of the relaxation time is smooth going from the sol phase to the transient gel phase. This question is currently under investigation and will be addressed in a forthcoming paper.

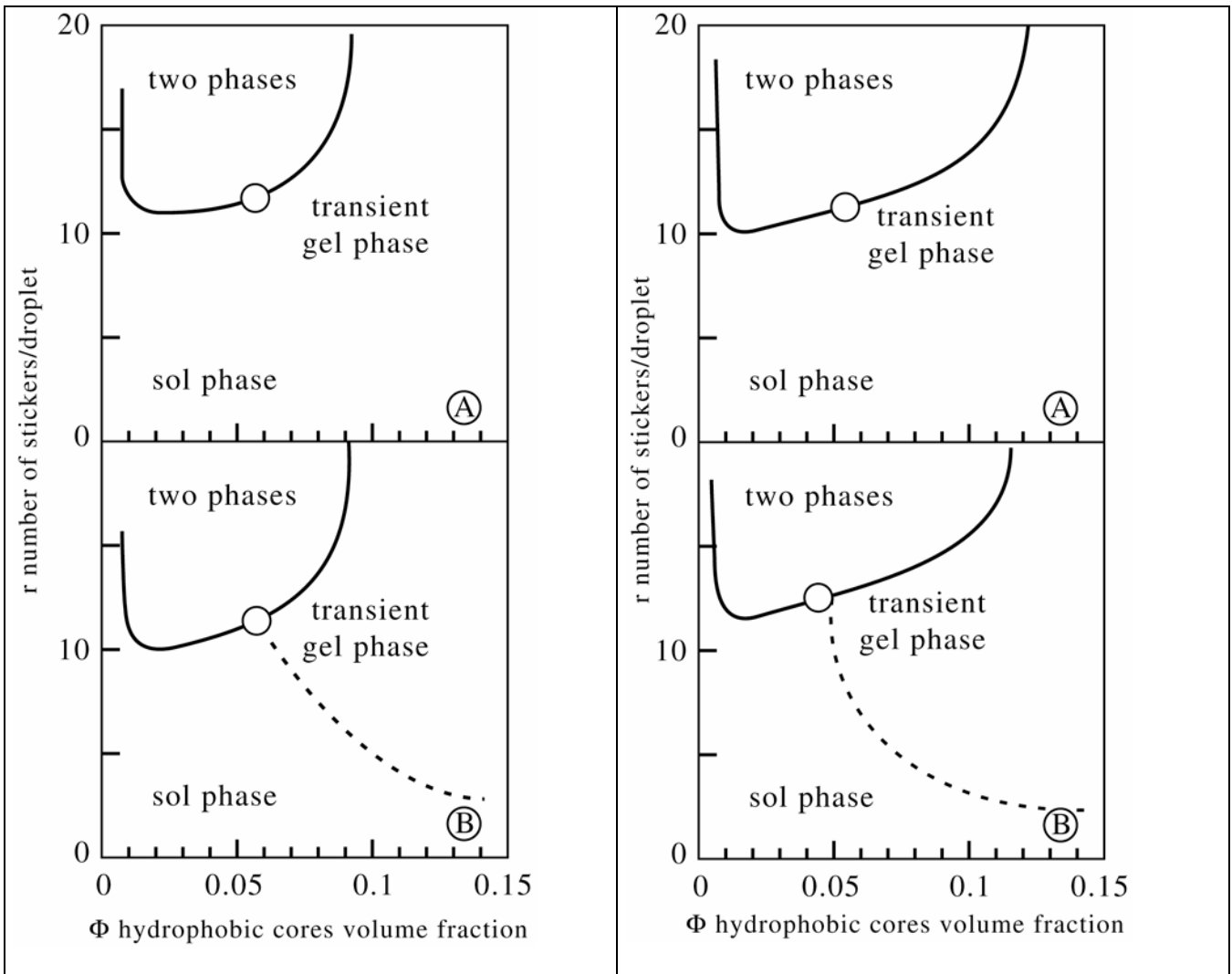

Figure 8 : Phase Behavior of the CPCl microemulsion A: upon addition of PEO-C12 at T= 20 °C and B: upon addition of PEO-C18 at T=

Figure 9 : Phase Behavior of the TX microemulsion A: upon addition of PEO-C12 at T= 25 °C and B: upon addition of PEO-C18 at T=

| 20 °C. Note the critical point = O associated to the phase separation and the percolation line ------ a tentative limit between the sol and the gel phase (see text). |  |
|---|---|

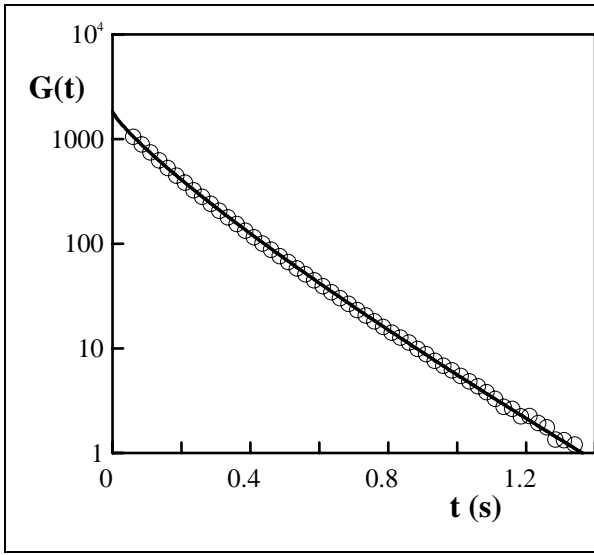

Figure 10

Stress relaxation curve after a step strain of $\gamma$= 20% for the sample $\Phi$ = 12.4 % and r= 13 of the TX microemulsion +PEO-C18 system. The solid line is the fit of the experimental data $G(t)= G(0) \exp(-t/\tau_R)^{0.82}$ with $G(0)$ =1830 Pa and $\tau_R$ =0.125 s

At low volume fraction and for moderate values of r (~> 10) a phase separation is observed between a very dilute sol phase and a concentrated gel phase. It has been described[6] for the CPCl / PEO-C12 system. The origin of the effective attraction responsible for this phase separation is discussed in the introduction and evidence for the existence of this effective attraction is found in the neutron scattering spectra of monophasic samples at low volume fraction and r < 10 (see above).

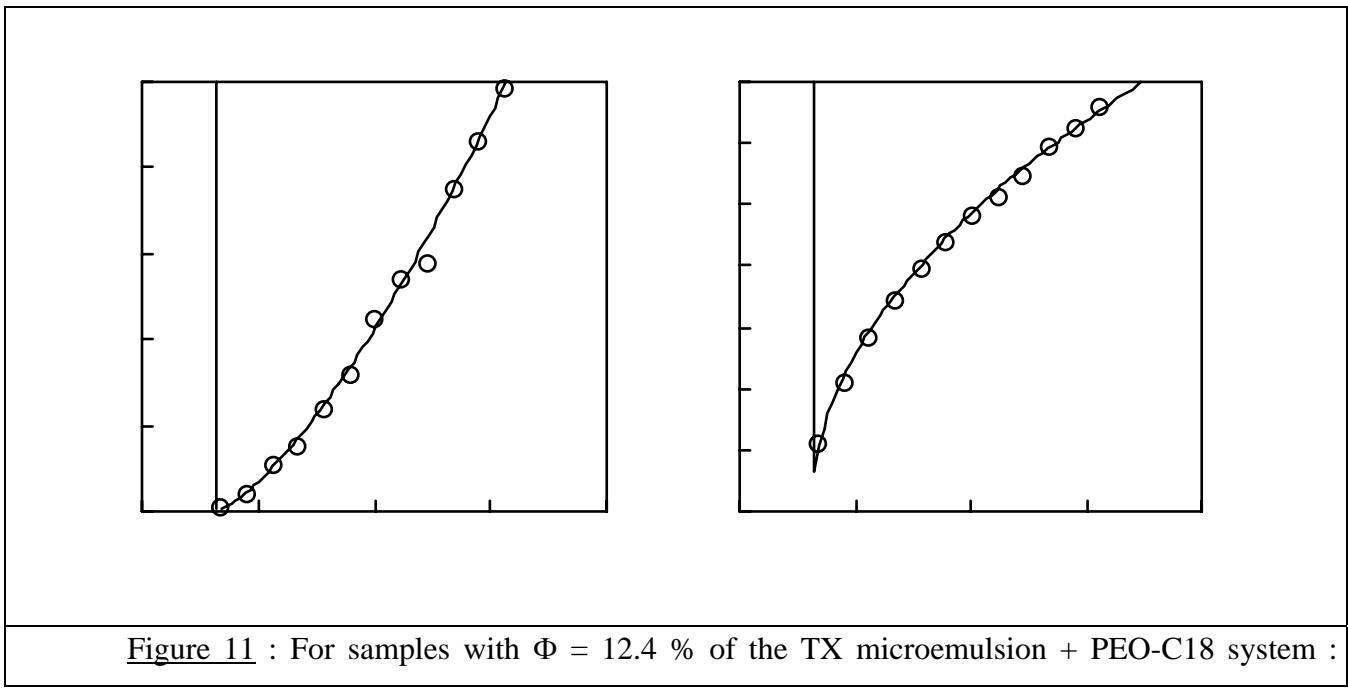

Figure 11 : For samples with $\Phi$ = 12.4 % of the TX microemulsion + PEO-C18 system :

> Evolution with r of G(0) and $\tau_R$ derived from the experimental results (as illustrated in figure 10 for the sample with r=13). <u>11A</u>: G(0), the solid line is the fit of the data to a power law G(0) = 68 (r-3.1)$^{1.42}$ Pa ; <u>11B</u> : $\tau_R$ the solid line is the fit of the data to a power law given by $\tau_R$ = 0.037 (r-3.1)$^{0.49}$ s.

Comparison of figure 8A or 8B for the CPCl system and of figure 9A or 9B for the TX system shows that, in each system, addition of PEO-C12 or PEO-C18 leads to the same Φ and r range (within experimental errors) for the phase separation. This observation is in good agreement with the description given for the effective attraction which must be independent of the nature of the stickers as long as the time they spend out of the droplets is negligible; it is easy to check that this is the case with $C_{18}$ and $C_{12}$ systems where the free energy variation going from a polar to an apolar surrounding i.e. the adhesion energy of the sticker can be estimated [9] to be in the order of 18 and 12 kT respectively. On the other hand the phase separation is observed in a similar but slightly different range of Φ and r in the CPCl or TX based systems. The main difference is the extension of the two-phase region to a larger volume fraction in the TX systems. This can be attributed to the fact that the overall interaction includes the interactions in the bare microemulsions, which are different as indicated above. The effective interaction introduced by the polymers is probably very similar but will depend on the distance between droplets. The distance between the hydrophobic surfaces of two droplets computed at the limit of the two-phase region in both systems is significantly different ~ 110 Å in the TX systems and ~ 90 Å in the CPCl systems. Such a difference can be partly due to the quality of the solvent (water or brine) for the PEO chain; water is a better solvent of PEO than brine: PEO swells more in the TX systems [32].

**Acknowledgments**

One of us, R.A., is grateful to Michel Viguier and André Collet for their precious advices during the synthesis of PEO-C12 and PEO-C18.

We thank Loic Auvray and Didier Lairez for their help during the SANS experiments performed on line PACE at Laboratoire Léon Brillouin -CEA-CNRS.

Table 1 Molar Mass and density of the components of the samples

| Component (abbreviated in the text) | Molar Mass (dalton) | HC(a) | Density (g/cm$^3$) polar part | HC(a) |
|---|---|---|---|---|
| $H_2O$ | 18 | - | 1 | - |
| $D_2O$ | 20 | - | 1.105 | - |
| $[H_3C-(CH_2)_{15}]-C_5H_5N^+ Cl^-$ (CPCl) | 339.5 | 225 | 1.656 | 0.83 |
| $[H_3C-(CH_2)_7]-OH$ (octanol) | 130 | 113 | 1.18 | 0.785 |
| $[H_3C-(C-(CH_3)_2-CH_2-C-(CH_3)_2)\varphi] (O-CH_2-CH_2)_{9.5}-OH$ (TX100) | 624 | 189 | 1.2 | 0.86 |
| $[H_3C-(C-(CH_3)_2-CH_2-C-(CH_3)_2)\varphi](O-CH_2-CH_2)_3-OH$ (TX35) | 338 | 189 | 1.2 | 0.86 |
| $[H_3C-(CH_2)_8 CH_3]$ (decane) | 142 | 142 | - | 0.75 |
| $[D_3C-(CD_2)_8 CD_3]$ (D-decane) | 164 | 164 | - | 0.86 |
| $[CH_3-(CH_2)_{11}]-NH-CO-(O-CH_2-CH_2)_{227}-O-(CO)-NH-[(CH_2)_{11} CH_3]$. (PEO-C12) | ~10 400 | 338 | 1.2 | 0.81 |
| $[CH_3-(CH_2)_{17}]-NH-CO-(O-CH_2-CH_2)_{227}-O-(CO)-NH-[(CH_2)_{17}CH_3]$. (PEO-C18) | ~10 600 | 506 | 1.2 | 0.81 |

a) HC = hydrophobic part of the component indicated in brackets in column 1

Table 2 Composition of the microemulsion droplets

| System | Components A and B of the surfactant layer (a) | Preparation in deuterated Water or Brine | | Preparation in Water or Brine | |
|---|---|---|---|---|---|
| | | $\Omega$ = A/B | $\Gamma$ = Decane/(A+B) | $\Omega$ = A/B | $\Gamma$ = Decane/(A+B) |
| CPCl | CPCl and Octanol | 0.25 | 0.62 | 0.25 | 0.56 |
| TX | TX100 and TX35 | 0.48 | 0.76 | 0.5 | 0.7 |

(a) The ratio of columns 2 to 4 are weight ratio

(b) Samples in deuterated water or brine have been prepared for the SANS measurements. For the determination of the phase behavior described in this paper and for the measurements of dynamical properties described elsewhere the samples are prepared in water or brine and it was found necessary to adjust slightly $\Omega$ and $\Gamma$ in order to remain close to the line of emulsification failure. We checked that the phase behavior was identical in both cases.

(32 )  The authors are grateful to the second refere for drawing their attention to this possibility.